\documentclass[a4paper,12pt]{article}
\usepackage{a4wide}
\usepackage[english]{babel}
\usepackage{amssymb, amsmath}

\newcommand{\bC}{{\bf C}}
\newcommand{\bg}{{\bf g}}

\newcommand{\bn}{\widehat{\bf n}}
\newcommand{\bOmega}{\boldsymbol{\Omega}}
\newcommand{\bV}{{\bf V}}
\newcommand{\bv}{{\bf v}}
\newcommand{\bw}{{\bf w}}
\newcommand{\kB}{k_{\scriptscriptstyle B}}
\newcommand{\Teff}{T_{\scriptscriptstyle eff}}
\newcommand{\ratio}{\mu}
\newcommand{\Etot}{E_{tot}}
\newcommand{\thetaeff}{\theta_{\scriptscriptstyle eff}}

\title{GRANULAR ROUGH SPHERE IN A LOW-DENSITY THERMAL BATH }
\author{F. Cornu \\
	Laboratoire de Physique Th\'eorique 
	\thanks{Laboratoire associ\'e au Centre National
	de la Recherche Scientifique - UMR 8627}\\
        B\^{a}timent 210, Universit\'e
	Paris-Sud, 91405 Orsay, France
	\and
	J. Piasecki\\ Institute of Theoretical Physics\\ University of Warsaw,
	Ho\.{z}a 69, 00-681 Warsaw, Poland}
\date{March 8, 2008}

\begin{document}

\maketitle

\begin{abstract}

We study the stationary state of a rough granular sphere immersed in a thermal bath composed of point particles. When the center of mass of the sphere is fixed the stationary angular velocity distribution is shown to be Gaussian with an effective temperature lower than that of the bath. For a freely moving rough sphere coupled to the thermostat via inelastic collisions we find a condition under which the joint distribution of the translational and rotational velocities  is a product of Gaussian distributions with the same effective temperature.  In this rather unexpected case we derive a formula for the stationary energy flow from the thermostat to  the sphere  in accordance with  Fourier law. 

\vskip 0.5cm 
{\bf PACS ~:} 45.70.-n, 05.20.Dd
\vskip 0.5cm 
{\bf KEYWORDS~:} tracer particle, rough granular sphere,  stationary state, Boltzmann's equation \vskip 0.5cm 
{\it Corresponding author ~:} CORNU Fran\c coise, \\
Fax: 33 1 69 15 82 87,
E-mail: francoise.cornu@u-psud.fr
\end{abstract}

\clearpage
\section{Introduction}

   The study of the dynamics of a tracer particle is a classical problem of nonequilibrium statistical mechanics. The object  is to determine the evolution of the state of a single particle resulting from interaction with the surrounding medium.  Such a study can yield precious information on the effects of many-body dynamics. Moreover, the relative simplicity of the problem creates chances for precise theoretical predictions. It has recently motivated  a number of works aiming at understanding the dynamics of fluidized granular media
\cite{BrilliantovPoeschel2004}. In particular, the evolution of a granular sphere immersed in a granular medium homogeneously cooling down has been discussed \cite{SantosDufty2006} as well as the Brownian motion in a granular fluid \cite{DuftyBrey2005}. The case of an impurity put in a vibrated low-density granular system has also been studied \cite{BreyRuiz-MonteroMoreno2006}.

An interesting qualitative question related to the effects of inelastic collisions taking place in granular fluids is that of the resulting structure of the distribution function when different kinds of degrees of freedom are present. A  gas of rough spheres with both translational and rotational degrees of freedom has been recently examined from this point of view \cite{BrilliantovPoeschelKranzZippelius2007}.  The main prediction based on numerical studies and approximate analytic arguments  is that dissipative collisions induce statistical dependence between orientations of the angular and translational velocities.

In the present paper we address an analogous question in an even simpler situation of a single tracer granular rough sphere suffering inelastic collisions with  point masses forming a low-density thermal bath. Our object is to find out  what kind of stationary state can result from a dissipative coupling to a thermostat. In the case of elastic collisions the particle would eventually attain equilibrium at the temperature of the bath. But the nature of the asymptotic stationary state in which there is a constant dissipative heat flow from the  thermostat to the tracer particle remains a largely open question of fundamental interest.

Exact results have been derived for a smooth hard sphere where  inelastic collisions could influence only  translational motion. It turned out that at the level of the Boltzmann kinetic theory the stationary velocity distribution had the form of a Maxwell distribution with an effective temperature lower than that of the thermostat
\cite{MartinPiasecki1999}. In one dimension one could even rigorously solve the initial value problem deriving in particular the exact dependence of the diffusion coefficient on dissipation \cite{PiaseckiTalbotViot2007}. In the case of  purely translational degrees of  freedom  the appearance of a Gaussian distribution has been shown to follow from the equivalence between the Boltzmann equation for a granular tracer particle suffering inelastic collisions and the Boltzmann equation for an elastic tracer particle with a suitably modified mass
 \cite{SantosDufty2006}.  

This remarkable property also occurs  if the test particle is rough but has a fixed mass center and thus only rotational degrees of freedom  \cite{Napiorkowski2006} (an original derivation of this fact is presented in Section 3). The stationary angular velocity is then again Gaussian with an effective temperature lower than that of the thermostat.

In the present paper we extend the study of stationary states at the level of Boltzmann's kinetic theory to the case of a tracer rough sphere whose both translational and rotational motions are influenced by inelastic collisions (the distribution of kinetic energy in a granular gas composed of rough spheres has been discussed in \cite{HuthmannZippelius1997}, \cite{LudingHuthmannMcNamaraZippelius1998} ). 
In section 2 we describe the model. Section 3  contains the description of our method first illustrated on simple situations where only one type of degrees of freedom is present. 
We then turn to the general case and show that when the restitution coefficients for the normal and tangential relative velocities obey a specific relation (\ref{constraint}) the joint velocity distribution becomes a product of two Maxwell distributions for the angular and translational velocities corresponding to the \textit{same} effective temperature. 
It turns out that the derived relation may be fulfilled only if the restitution coefficient relevant for rotational motion is larger than that linked to the motion of the mass center. 

The occurence of a stationary factorized Gaussian distribution for the two types of degrees of freedom  inelastically excited by collisions is quite remarkable, and, in view of the results obtained for a gas of granular rough spheres \cite{BrilliantovPoeschelKranzZippelius2007} rather unexpected. 
The main result of Section 3 is obtained by using an appropriate change of integration variables in the gain term of the Boltzmann equation ( the method generalizes that used in Ref. \cite{SantosDufty2006} ).

The heat flux that maintains the test particle in a stationary state is calculated in Section 4. It obeys the analogue of Fourier  law with a thermal conductivity proportional to the temperature jump between the sphere and the thermostat, as in  generic hydrodynamic theories. 
In Section 5 we briefly comment on the possible structure of the joint velocity distribution when the values of the restitution coefficients are not related by the equation derived in Section 3. We expect that the typical case would involve statistical dependence between the angular and translational velocities of the sphere.

\section{The model}

For the sake of simplicity the thermal bath particles are supposed to be point masses $m$ performing purely translational motion.
Their distribution  in the one-particle phase space is the product of a uniform spatial density 
$\rho$ and a Maxwell velocity distribution $\phi_T(\bv;m)$ corresponding to temperature $T$
\begin{equation}
\label{Maxwellian}
\phi_T(\bv;m)=\left(\frac{m}{2\pi\kB T}\right)^{D/2} \exp\left[-\frac{m\bv^2}{2\kB T} \right]
\end{equation}
$D$ is the dimension of the space ($D=2$
 or $3$), and $k_{B}$ is  Boltzmann's constant.

The rough sphere is supposed to have mass  $M$, radius $R$, moment of  inertia $I$, and to move with  translational velocity $\bV$,  and angular velocity $\bOmega$.
Its total kinetic energy equals thus
\begin{equation}
\label{defE}
E(\bV, \bOmega) = \frac{1}{2}M\bV^2+\frac{1}{2} q M R^2 \bOmega^2
\end{equation}
where $q=I/MR^2$ is a number reflecting the mass density distribution inside the sphere (disk).

\subsection{Collisional laws}

Consider a binary collision between the rough sphere and a point particle of the thermostat. The instantaneous collisional  transformation of velocities 
\begin{equation}
( \bV, \bOmega, \bv )  \rightarrow ( \bV^*, \bOmega^*, \bv^* ) 
\label{dcollision}
\end{equation}
is conveniently described with the help of the unit vector $\bn$  along the line segment from  the center of the sphere to the point of impact.  

The linear velocity of the point at the surface of the sphere hit by the thermostat particle is  
$( \bV+ R \bOmega\times \bn )$. 
The relative velocity at which the particle approaches the impact point is thus
\begin{equation}
\label{defbg}
\bg=\bv-\bV-R\bOmega\times \bn
\end{equation}
In what follows we will use the notations 
\begin{equation}
{\bf A}_n=\left({\bf A}\cdot\bn\right)\bn, \;\;\;\;\;
{\rm and} \;\;\;\;\;
{\bf A}_t={\bf A}-{\bf A}_n=\bn\times\left({\bf A}\times\bn\right)
\end{equation}
for the  normal and tangential components of any vector ${\bf A} $. 

In the simplest model of inelastic collisions one defines the instantaneous transformation of relative velocity (\ref{defbg}) by (see e.g.\cite{BrilliantovPoeschel2004})
 \begin{equation}
 \label{alpha}
 \bg_n^{\star}=-\alpha\bg_n
\end{equation}
\begin{equation}
\label{beta}
\bg_t^{\star}=-\beta\bg_t
\end{equation}
where $0\leq \alpha\leq 1$, and  $-1\leq \beta\leq 1$ are the translational and rotational restitution coefficients, respectively. The cases of $\alpha =1$ with $\beta =\pm1$ correspond to elastic encounters. 

In order to calculate postcollisional velocities $ ( \bV^*, \bOmega^*, \bv^* )$ one has  to use relations 
(\ref{alpha}),(\ref{beta}) together with the conservation laws for both the momentum and the angular momentum.
One then finds (see e.g. \cite{LudingHuthmannMcNamaraZippelius1998})
\begin{eqnarray}
\label{velocitystar}
R \bOmega^{\star}&=&R \bOmega+ \frac{1+\beta}{1+ q(1+\ratio)}\bn\times \bg_t
\nonumber\\
\bV^{\star}&=&\bV+\frac{1+\alpha}{1+\ratio}\bg_n +\frac{q(1+\beta)}{1+ q(1+\ratio)}\bg_t
\\
\bv^{\star}&=&\bv-\frac{\ratio(1+\alpha)}{1+\ratio}\bg_n - 
\frac{q\ratio (1+\beta )}{1+ q(1+\ratio)}\bg_t
\nonumber
\end{eqnarray}
where $\ratio$ denotes the mass ratio  $\ratio = M/m$.

According to the defining equations (\ref{alpha}) and (\ref{beta}) 
the formulae corresponding to the inverse collision
\begin{equation}
( \bV^{\star\star}, \bOmega^{\star\star}, \bv^{\star\star} )  \rightarrow ( \bV, \bOmega, \bv ) 
\label{icollision}
\end{equation}
are obtained   by changing $\alpha$ into $1/\alpha$, and $\beta$ into $1/\beta$.

\subsection{The Boltzmann equation}

We denote by $f(\bV,\bOmega,t)$ the probability density  for finding the test particle at time $t$ with translational velocity $\bV$ and angular velocity $\bOmega$. 

Within Boltzmann's theory whose predictions are the object of the
present paper, the tracer particle never suffers recollisions, and thus always encounters unperturbed thermalized particles. The Boltzmann kinetic equation is thus  linear in this case and in homogeneous situations it has the form
\begin{eqnarray}
\label{eqBbis}
\frac{\partial f(\bV,\bOmega,t)}{\partial t}
&=&\rho R^{D-1} \int d\bn \int d\bv \,
\Theta\left(\left[\bv-\bV\right]\cdot \bn\right)\vert\left[\bv-\bV\right]\cdot \bn\vert
\\
&&\qquad\qquad\times \left[\frac{1}{\alpha^2\vert\beta\vert^{D-1}}
 f(\bV^{\star\star},\bOmega^{\star\star},t)\phi_T(\bv^{\star\star};m) 
- f(\bV,\bOmega,t)\phi_T(\bv;m)\right] 
\nonumber
\end{eqnarray}
 In the gain term there appears the factor $1/\alpha^2\vert\beta\vert^{D-1}$ which guarantees the conservation of the normalization of the velocity distribution. $\Theta(x)$ denotes here the unit step function. We restrict further discussion to the stationary solution of equation (\ref{eqBbis}).

 It will be useful to display the detailed structure of the gain term. To this end we express the precollisional velocities 
$( \bV^{\star\star}, \bOmega^{\star\star}, \bv^{\star\star} ) $ in terms of the scaled relative velocities
\begin{equation}
\label{scaledrvn}
\bw_n=\frac{1}{\alpha}\bg_n= \frac{1}{\alpha}\left(\bv-\bV\right)_n
\end{equation}
\begin{equation}
\label{scaledrvt}
\bw_t=\frac{1}{\beta} \bg_t=\frac{1}{\beta}\left(\bv-\bV - R \bOmega \times \bn\right)_t
\end{equation}
Using equations (\ref{velocitystar}) with $\alpha$ and $\beta$ replaced by $\alpha^{-1}$ and 
$\beta^{-1}$  we find
\begin{eqnarray}
\label{change}
R\bOmega^{\star\star} & = &R \bOmega - \frac{(1+\beta)}{1+q(1+\ratio)}\bw_t \times \bn
\nonumber 
 \\
\bV^{\star\star} & = & \bV + \frac{1+\alpha}{1+\ratio}\bw_n + \frac{q(1+\beta)}{1+q(1+\ratio)}\bw_t  \\
\bv^{\star\star} & = & \bV + R \bOmega\times\bn - \bw +  \frac{1+\alpha}{1+\ratio}\bw_n +
\frac{(1+q)(1+\beta)}{1+q(1+\ratio)}\bw_t \nonumber
\end{eqnarray}
Under the change of the integration variables $(\bv_n , \bv_t ) \rightarrow (\bw_n , \bw_t )$  the gain term takes the form
\begin{eqnarray}
\label{gaincorrige}
G(\bV ,\bOmega) &=&
 \rho R^{D-1} \int d\hat{\bf n}\int_{0}^{\infty}dw_{n} w_{n}\int d\bw_{t} 
 \\
&&\qquad f\left( \bV + \frac{1+\alpha }{1+\ratio }\bw_{n} + \frac{q(1+\beta )}{1+q(1+\ratio )} \bw_{t}, \; \bOmega -
\frac{(1+\beta)(\bw_{t}\times\hat{\bf n})}{R[1+q(1+\ratio)]} \right) 
\nonumber\\
 &&\qquad \times \phi_T\left(  \bV + R\bOmega\times\hat{\bf n} - \bw +\frac{1+\alpha }{1+\ratio }\bw_{n}
+ \frac{(1+q)(1+\beta)}{1+q(1+\ratio)}\bw_{t} ;m\right) 
\nonumber
\end{eqnarray}
where $ w_{n}= |\bw_{n}| $, and the relation $d\bv_n d\bv_t  =  \alpha\vert\beta\vert^{D-1}d\bw_n d\bw_t$ has been taken into account. The use of the scaled relative velocities (\ref{scaledrvn}), (\ref{scaledrvt}) as integration variables will play a crucial role in exploring the possibility of finding an elastic problem equivalent to the dissipative one.

\subsection{Energy dissipated in a binary collision}
 
The collisional transformation laws (\ref{velocitystar}) 
conserve the  center of mass velocity
\[ \bV_G=\frac{M\bV+m\bv }{M+m} = \frac{\ratio\bV + \bv}{\ratio + 1}, \]
the tangential vector
\begin{equation}
\label{Ct}
\bC_t=\bv_t-\bV_t+ q(1+\ratio)R\bOmega_t\times\bn ,
\end{equation}
and also the normal component of the angular velocity $\bOmega_n$. Moreover, the transformation laws for the relative velocity $\bg$ reduce to simple rescalings (\ref{alpha}), (\ref{beta}). 

The evaluation of the collisional change in the total kinetic energy of the colliding pair
\[ \Delta \Etot \equiv \Etot(\bV^{\star},\bOmega^{\star},\bv^{\star})-\Etot(\bV,\bOmega,\bv) \]
where
\begin{equation}
\label{Etot}
\Etot (\bV,\bOmega,\bv) = E(\bV,\bOmega) + \frac{1}{2}m\bv^2 =
 \frac{1}{2}m\left\{\ratio \bV^2+\frac{1}{2}q\ratio R^2\bOmega^2 + {\bv^2}\right\},
\end{equation}
can be thus conveniently performed by expressing $\Etot$  in terms of the  invariant vectors
$\bV_G$, $\bOmega_n$, $\bC_t$, and the relative velocity $\bg$. 
Owing  to the relations
\begin{equation}
%\label{}
\bv^2+\ratio \bV^2=(1+\ratio)\bV_G^2+ \frac{\ratio}{1+\ratio}\left(\bv-\bV\right)^2
\end{equation}
and
\begin{equation}
\left(\bv-\bV\right)_t^2+ q(1+\ratio)R^2\bOmega_t^2= 
\frac{1}{1+q(1+\ratio)}\left[q(1+\ratio)\bg_t^2 +\bC_t^2\right]
\end{equation}
we find 
\begin{equation}
%\label{}
\Etot= \frac{m}{2} \frac{\ratio}{(1+\ratio )}
\left\{\frac{(1+\ratio)^2}{\ratio }\bV_G^2
+\left[ \bg_n^2 +q(1+\ratio) R^2\bOmega_n^2\right]
+\frac{1}{1+q(1+\ratio)}\left[ q(1+\ratio)\bg_t^2 + \bC_t^2\right]\right\}
\end{equation}
The dissipation of the total energy  $\Etot$ under binary collisions is due to the reduction of the length of vectors $\bg_n$ and $\bg_t$ governed by the restitution coefficients $\alpha$ and $\beta$. In view of the invariance of vectors $\bV_G$, $\bOmega_n$, and $\bC_t$ one finds
 \begin{equation}
\label{DeltaEtot}
\Delta \Etot  =- \frac{m}{2}\frac{\ratio}{(1+\ratio )}
\left\{(1-\alpha^2)\bg_n^2+ \frac{q(1+\ratio)}{1+q(1+\ratio)}(1-\beta^2)\bg_t^2\right\}
\end{equation}
The total pair energy is conserved if $\alpha=1$ and $\beta=\pm 1$.
 
\clearpage 
\section{Solvable situations}

\subsection{Specific elastic cases}
 
In the case of an elastic  rough sphere  $\alpha = \beta=1$, the relative velocity $\bg$ is reversed as the result of the impact, and the conservation of the kinetic energy $\Delta \Etot =0$ implies the equilibrium solution of the Boltzmann equation involving the product of two Gaussian distributions  
\begin{equation}
\label{prodMaxwellian}
f^{eq}(\bV,\bOmega)=\phi_T(\bV;M)\phi_T^{rot}(\bOmega;I)
\end{equation}
satisfying the equipartition law. Here
\begin{equation}
\phi_T^{rot}(\bOmega;I)=\left(\frac{I}{2\pi\kB T}\right)^{D/2}
\exp\left(-\frac{I\bOmega^2}{2 \kB T}\right)
\end{equation}
with $I$ defined in Eq.(\ref{defE}).
The Maxwell distribution $\phi_T(\bV;M)$ has been defined in Eq.(\ref{Maxwellian}).

The case of $\alpha =1$, $ \beta=- 1$ corresponds to an elastic smooth sphere: the rotational degrees of freedom are not influenced by collisions. Consequently,  any distribution of the form
\begin{equation}
\phi_T(\bV;M)\chi (\bOmega)
\end{equation}
represents a stationary state of a smooth sphere. 

\subsection{Connection with an elastic problem for a rough sphere}

Let us consider now the possibility for reducing the Boltzmann equation for a granular sphere (inelastic collisions) to an equivalent case of a rough sphere elastically coupled to the thermostat. The main observation 
is that the gain term (\ref{gaincorrige}) would take the elastic form for a rough sphere ($\alpha=1$ and $\beta=1$) with a mass $\widetilde{\ratio} m$
\begin{eqnarray}
\label{gainelcorrige}
&&\rho R^{D-1} \int d\hat{\bf n}\int_{0}^{\infty}dw_{n} w_{n}\int d\bw_{t} 
f\left( \bV + \frac{2 }{1+\widetilde{\ratio }}\bw_{n} + \frac{2q}{1+q(1+\widetilde{\ratio})} \bw_{t}, \bOmega -
\frac{2(\bw_{t}\times\hat{\bf n})}{R[1+q(1+\widetilde{\ratio})]} \right)
\nonumber\\
&&\qquad\qquad
\times \phi_T\left(  \bV + R\bOmega\times\hat{\bf n} - \bw + \frac{2 }{1+\widetilde{\ratio }}\bw_{n}
+ \frac{2(1+q)}{1+q(1+\widetilde{\ratio})}\bw_{t} ;m\right) 
\end{eqnarray}
provided the equations
\begin{equation}
\label{conditions} 
\frac{1+\alpha}{1+\ratio}  =   \frac{2}{1+\widetilde{\ratio}} \;\;\;\;\;\; {\rm and} \;\;\;\;\;\;\;
\frac{1+\beta }{1+q(1+\ratio )}  =  \frac{2}{1+q(1+\widetilde{\ratio })}   
\end{equation}
could be simultaneously satisfied. This simple fact is the basis for the further discussion. The corresponding formula for the loss term 
\begin{equation}
\label{Lbis}
L(\bV ,\bOmega)=\rho R^{D-1}\int d\bn\int_{0}^{+\infty} d g_n g_n \int d\bg_t
 f(\bV,\bOmega)\phi_T\left(\bV + R\bOmega \times \bn+\bg;m\right)
\end{equation}
is irrelevant from the point of view of the correspondence with an elastic case as it does not involve the collision law.

Before discussing the general case let us consider two simple situations where only one type of degrees of freedom is present.

When $\beta =-1$, the sphere is smooth and only the translational velocity $\bV$ is influenced by collisions.  We are thus interested in the stationary velocity distribution $f_{st}(\bV)$.  For a comparison with  a smooth sphere undergoing elastic collisions with the point particles of the thermostat, we make the extra change of variable 
$\bw_t  \rightarrow -\bw_t$ in the gain term (\ref{gaincorrige}) which becomes for a smooth sphere ($\beta=-1$)
\begin{eqnarray}
\label{Glisse}
G(\bV,\bOmega) &= &\chi(\bOmega) 
 \rho R^{D-1} \int d\hat{\bf n}\int_{0}^{\infty}dw_{n} w_{n}\int d\bw_{t} 
 \\
 &&\qquad\qquad f\left( \bV + \frac{1+\alpha }{1+\ratio }\bw_{n} \right) 
  \phi_T\left(  \bV + R\bOmega\times\hat{\bf n} + \bw -\left(2-\frac{1+\alpha }{1+\ratio }\right)\bw_{n} ;m\right) 
 \nonumber
\end{eqnarray}
(The complete variable change is similar to that used in Ref. \cite{SantosDufty2006}.)
The gain term (\ref{Glisse}) would take the elastic form  for a smooth sphere ($\alpha=1$ and $\beta=-1$) with a mass $\widetilde{\ratio} m$
\begin{eqnarray}
\label{gainellisse}
&&\chi(\bOmega) \rho R^{D-1} \int d\hat{\bf n}\int_{0}^{\infty}dw_{n} w_{n}\int d\bw_{t} 
 \\
 &&\qquad\qquad\qquad
f\left( \bV + \frac{2 }{1+\widetilde{\ratio }}\bw_{n} \right)
 \phi_T\left( \bV + R\bOmega \times \bn+\bw - \frac{2 \widetilde{\ratio } }{1+\widetilde{\ratio }}\bw_{n} ;m\right) 
 \nonumber
\end{eqnarray}
provided that $(1+\alpha)/(1+\ratio)=2/(1+\widetilde{\ratio})$.
Eventually, from the two conditions 
(\ref{conditions}) only the first remains relevant yielding the mass ratio 
\begin{equation}
\label{valuewidetilderatio1}
\widetilde{\ratio}=\ratio +\frac{1-\alpha}{1+\alpha}(1+\ratio)
\end{equation}
The stationary distribution of the mass center velocity has thus the form of the Maxwell distribution at temperature $T$ for a particle with  mass $\widetilde{M}=\widetilde{\ratio} m$,
or, equivalently, it is a Maxwell distribution for the sphere of mass $M$ with an effective temperature $T_{*}$ satisfying the relation $\widetilde{M}/T = M/T_{*}$. From (\ref{valuewidetilderatio1}) we find
\begin{equation}
\label{Teff1}
T_{*}=\frac{(1+\alpha)\ratio}{1-\alpha+2\ratio}T\quad <\quad T
\end{equation}
Hence $f_{st}(\bV)=\phi_{T_{*}}(\bV; M)$
 which is the result derived in \cite{MartinPiasecki1999}.

When the rough sphere has a clamped center of mass, $\bV=\boldsymbol{0}$,  only rotational degrees of freedom are involved. The corresponding collision laws can be deduced from equations (\ref{velocitystar})
by taking the limit $\ratio\to\infty$ and  $q \to 0$,
while keeping the product $q\ratio= I/mR^{2}$  fixed. Indeed, in this limit the moment of inertia is kept unchanged while the mass relevant to translational motion tends to infinity which immobilizes the center of mass of the sphere. We are thus interested in the stationary velocity distribution $f_{st}(\bOmega)$.
The gain term (\ref{gaincorrige}) takes the form
\begin{eqnarray}
\label{gainrotcorrige}
&&G(\bOmega) =
 \rho R^{D-1} \int d\hat{\bf n}\int_{0}^{\infty}dw_{n} w_{n}\int d\bw_{t} 
 \\
&&\quad\quad\quad f\left(\bOmega - \frac{1+\beta}{R[1+q\ratio ]}\bw_{t}\times\hat{\bf n} \right) 
  \phi_T\left(  R\bOmega\times\hat{\bf n}  - \bw  + \frac{1+\beta}{1+q\ratio }\bw_{t} ;m\right)
 \nonumber
\end{eqnarray}
 The gain term (\ref{gainrotcorrige}) would take the elastic form  for a clamped rough sphere ($\bV=\boldsymbol{0}$, $\alpha=1$ and $\beta=1$) with a mass $\widetilde{\ratio} m$
 \begin{eqnarray}
&& \rho R^{D-1} \int d\hat{\bf n}\int_{0}^{\infty}dw_{n} w_{n}\int d\bw_{t} 
 \nonumber\\
&&\qquad\qquad
\times 
 f\left(\bOmega - \frac{2}{R[1+q\widetilde{\ratio} ]} \bw_{t}\times\hat{\bf n}\right) 
  \phi_T\left(  R\bOmega\times\hat{\bf n}  - \bw  + \frac{2}{1+q\widetilde{\ratio}}\bw_{t} ;m\right)
\end{eqnarray}
provided that 
\begin{equation}
\label{constraintrot}
 \frac{1+\beta }{1+ q\ratio}  =  \frac{2}{1+ q\widetilde{\ratio } }   
\end{equation} 
Eventually in the above mentioned limit only the second of the two relations (\ref{conditions}) remains  relevant : it takes the limit form (\ref{constraintrot})
and yields the mass ratio
\begin{equation}
\label{valuewidetilderatio2}
\widetilde{\ratio}=\frac{1-\beta+2q\ratio}{ (1+\beta) q\ratio }\ratio
\end{equation}
The stationary solution of the Boltzmann equation  has thus again the form of a Maxwell distribution for the angular velocity $\bOmega$
 \begin{equation}
\label{fstcenterfixed}
f_{st}(\bOmega)=
 \left(\frac{qMR^2}{2\pi T_{**}^0}\right)^{D/2}
\exp\left[-\frac{qM R^2\bOmega^2}{2 \kB T_{**}^0} \right]
\end{equation}
with the effective temperature $T_{**}^{\,0}$ given by
\begin{equation}
\label{Teff2}
T_{**}^{\,0}=\frac{(1+\beta)q\ratio}{1-\beta+2q\ratio } T \quad < \quad T
\end{equation}
The results (\ref{fstcenterfixed}),(\ref{Teff2}) have been previously derived in an unpublished work  
\cite{Napiorkowski2006} in a different way. We notice that $T_{**}^{\,0}$ is independent of $\alpha$, whereas the collisional laws are not.

In the general case of a granular rough sphere ($-1 < \beta \le +1 $, $0\le \alpha < 1$) with translational and rotational degrees of freedom both conditions (\ref{conditions}) are to be satisfied simultaneously implying the relation between the restitution coefficients of the form
\begin{equation}
\label{constraint}
\beta=\alpha+\frac{1-\alpha^2}{1+\alpha+2q(1+\ratio)} 
\end{equation}
The relation can occur only for  positive values of the restitution coefficient $\beta $ that are larger than $ \alpha $. 
It is thus possible to rewrite the gain term (\ref{gaincorrige}) in the form corresponding to elastic collisions  provided   equation (\ref{constraint}) is satisfied. This fact is quite remarkable, and, as we have mentioned in the introduction, rather unexpected.
It would be interesting to understand why the particular tuning (\ref{constraint}) of the parameters governing the dissipation makes appear the Gaussian distribution.

We notice that  the constraint (\ref{constraint}) can be rewritten as $T_{*}=T_{**}$ with $T_{*}$ given in (\ref{Teff1}) and $T_{**}$ defined by
\begin{equation}
\label{defTstarstar}
T_{**}=\frac{(1+\beta)q\mu}{(1+q)(1-\beta)+2q\mu} T
\end{equation}
The notations have been chosen in order to take into account the property 
\hfill\break $T_{**}^{\,0}=\lim_{q\to 0, \textrm{$q\mu$ fixed}}T_{**}$ (see (\ref{Teff2})). When
$T_{**}$  is expressed in terms of $T_{**}^{\,0}$, the condition $T_{*}=T_{**}$ takes the form
\begin{equation}
\label{relation}
T_*= \left[1+\frac{1-\beta}{\ratio(1+\beta)}\frac{T_{**}^{\,0}}{T}\right]^{-1}T_{**}^{\,0}
\end{equation}
where each temperature $T_*$ or $T_{**}^{\,0}$ depends only on one restitution coefficient, $\alpha$ or $\beta$  respectively.
In other words, the constraint
(\ref{constraint}) expresses the relation (\ref{relation}) between the effective temperatures found in the study of a sphere for which collisions could influence either  only translational or only rotational  motion.   
Our analysis shows that the gain term in the Boltmann equation can be rewritten in the form corresponding to elastic collisions only when  the dissipation related to pure translational motion and the dissipation related to pure rotations would lead to effective temperatures linked by equation (\ref{relation}). 
The stationary state of the granular rough sphere under the condition (\ref{constraint}) complies thus with the energy equipartition and reads
\begin{equation}
\label{fst}
f_{st}(\bV,\bOmega)=\left(\frac{M}{2\pi\Teff}\right)^{D/2}
 \left(\frac{qMR^2}{2\pi\Teff}\right)^{D/2}
\exp\left[-\frac{1}{\kB \Teff}
 \left(\frac{1}{2} M\bV^2+\frac{1}{2} qM R^2\bOmega^2\right)\right]
\end{equation}
with $\Teff = T_{*}=T_{**}$.

Clearly, the equivalence with the elastic problem on curve (\ref{constraint}) also
holds for the Boltzmann equation in the presence of an arbitrary
inhomogeneous nonequilibrium state of the medium surrounding the test
particle, since  the structure of the collision term in this general case
remains unchanged. It seems worth mentionning here that also the Enskog
equation could be reduced to its elastic form on curve (\ref{constraint}), since  it involves
exactly the same collisional transformation of velocities as the Boltzmann
equation (see e.g.\cite{SantosDufty2006}).

\clearpage
\section{Analogue of  Fourier law in  the solvable stationary state}

In the stationary state determined by the relation \eqref{constraint} 
the mean kinetic energy of the test particle results from the energy equipartition between the  rotational and translational degrees of freedom
\begin{equation}
\int d\bV\int d\bOmega f_{st}(\bV,\bOmega)
\left[\frac{1}{2}M\bV^2+\frac{1}{2} q M R^2 \bOmega^2\right]=D \kB \Teff
\end{equation}
($D$ is the  dimension of the position space.)
This stationary state is sustained by an energy flux from the heat bath to the test particle that compensates the energy dissipation under binary collisions.

The collisional rate of the energy dissipation equals
\begin{equation}
J_{E}^{diss}=\int d\bV\int d\bOmega\int d\bv\int d\bn \Theta\left(\left[\bV-\bv\right]\cdot\bn\right)
 \left[\bV-\bv\right]\cdot\bn \; \Delta \Etot \;f_{st}(\bV,\bOmega)\phi_T(\bv;m)
\end{equation}
Inserting here  $\Delta \Etot $ calculated in equation (\ref{DeltaEtot}) one finds  
\begin{eqnarray}
&& J_{E}^{diss} =
-\frac{D-1}{2}\sqrt{1+\frac{\thetaeff}{\ratio}}
\left\{
 \frac{2(\ratio+\thetaeff)}{1+\ratio}(1-\alpha^2)
+\frac{(D-1)[q\ratio+(1+q)\thetaeff]}{1+q + q\ratio}(1-\beta^2)\right\}
\nonumber
\\
&&\qquad\qquad \times
\kB T\sqrt{\frac{2\pi\kB T}{m}}
\end{eqnarray}
where $\thetaeff=\Teff/T$ and $\Teff$ is defined in (\ref{fst}).  Since $\alpha$ and $\beta$ obey the constraint \eqref{constraint},
the relations \eqref{Teff1} and \eqref{defTstarstar} allow one to express $\alpha$ and $\beta$ in terms of $\thetaeff$. In particular
\begin{equation}
1-\alpha^2=4 \frac{\ratio(1+\ratio)}{(\ratio +\thetaeff)^2}(1-\thetaeff)\thetaeff
\end{equation}
and
\begin{equation}
1-\beta^2=4 \frac{q\ratio[1+q+ q\ratio]}{[q \ratio+(1+q)\thetaeff]^2}(1-\thetaeff)
\thetaeff
\end{equation}
Eventually one finds that the heat flux   obeys the analogue of  Fourier law
\begin{equation}
J_{E}^{diss} = \kappa(T) \left(\Teff-T\right)
\end{equation}
in which the discontinuous jump of the temperature plays the role of the
temperature gradient present in the theory of continuous media.
The thermal conductivity for $D=2$ or $3$ reads
\begin{equation}
\kappa(T)= 2(D-1)\thetaeff \sqrt{1+\frac{\thetaeff}{\ratio}}
\left(\frac{2\ratio}{\ratio+\thetaeff}+\frac{(D-1)q\ratio}{q\ratio+(1+q) \thetaeff}\right)
\kB \sqrt{\frac{2\pi \kB T}{m}}
\end{equation}
where $\thetaeff$ is a dimensionless function of $\alpha$  and $\ratio$ given by \eqref{Teff1} or equivalently a function of 
$\beta$ and $q\ratio$  as given by \eqref{defTstarstar}.

\section{Open questions}

Outside the curve (\ref{constraint}) in the $(\alpha,\beta)$ plane the possibility of deriving the analytic form of the stationary velocity distribution remains an interesting open problem.
The simplest generalization of solution (\ref{fst}) would be the product of two Gaussian distributions with different effective temperatures for translational and rotational motions. However, on the basis of detailed calculations we conjecture that such a distribution is not possible, and that the Gaussian form can appear only on the derived curve (\ref{constraint}). The complete proof of the conjecture is still to be constructed.

Since the translational velocity $\bV$ is a vector whereas the angular velocity $\bOmega$ is a pseudovector, 
the stationary state can depend in general on three scalar variables $\vert\bV\vert$, $\vert\bOmega\vert$ and $\vert\widehat{\bV}\cdot\widehat{\bOmega}\vert$, where $\widehat{\bV}$ and $\widehat{\bOmega}$ are the unit vectors in the direction of $\bV$ and $\bOmega$ respectively. We conjecture that when the constraint (\ref{constraint}) is not respected the stationary distribution depends on the angle between $\bV$ and $\bOmega$ via the variable $\vert\widehat{\bV}\cdot\widehat{\bOmega}\vert$  which introduces statistical dependence between the two velocities. 
\medskip   

{\bf Acknowledgments}

J. P.  acknowledges the hospitality at the Laboratoire de Physique Th\'eorique de
l'Universit\'e Paris-Sud, and the Ministry of Science and Higher Education (Poland) for partial financial support (research project No. N20207631/0108).

\end{document}